\documentstyle[aps,preprint]{revtex}

\def \beq {\begin{equation}}
\def \eeq {\end{equation}}

\begin{document}

\draft

\title{Nonexistence theorems for traversable wormholes}

\author{Alberto Saa}
\address{
Institut f\"ur Theoretische Physik,\\
Freie Universit\"at Berlin, Arnimallee 14,
14195 Berlin, Germany \\ and \\
Departamento de Matem\'atica Aplicada, \\ IMECC--UNICAMP,
        C.P. 6065, 13081-970 Campinas, SP, Brazil}
\maketitle

\begin{abstract}
Gauss--Bonnet formula is used to derive a new and 
simple theorem of nonexistence of
vacuum static nonsingular lorentzian wormholes. 
We also derive simple proofs for
 the nonexistence of lorentzian wormhole solutions
for some classes of
static matter such as, for instance, real scalar fields 
with a generic potential obeying $\phi V'(\phi) \ge 0$ and
massless fermions fields.
\end{abstract}

\newpage

A lorentzian wormhole is a solution of Einstein equations with asymptotically
flat regions connected by intermediary ``throats''. These solutions
have received considerable attention since Morris,
Thorne and Yurtsever\cite{w1,w2} discussed the possibility of 
traversing them and their connection with time machines. 
(For a review, see \cite{viss}.) 
Euclidean
wormholes, {\em i.e.}, wormhole solutions of Einstein equations with
signature (+,+,+,+), have also been intensively studied in connection
with the cosmological constant problem\cite{cc}.

Originally, the analysis of lorentzian wormholes was restricted
to the static spherically symmetrical case\cite{w1,w2}. 
Birkhoff's theorem assures
that the only vacuum spherically symmetrical lorentzian wormhole
is the maximally extended Schwarzschild solution. However, in this case
the ``throat''connecting the two asymptotically flat regions
($r\rightarrow\pm\infty$) is singular for $r=0$, and hence it is
not traversable\cite{w1,w2,mtw}. 
We recall that a lorentzian wormhole solution is said to be
traversable\cite{w1,w2} if it does not contain horizons that prevent
the crossing of the ``throats'' and if an observer crossing them
does not experience strong tidal forces.
In order to obtain a solution with a traversable
wormhole, we are enforced to accept the presence of matter and/or to
give up of the spherical symmetry. A list of recent solutions includes
stationary\cite{teo} and static axisymmetric electrovac\cite{SA} cases, and
solutions in
Brans-Dicke\cite{BD},  Kaluza-Klein\cite{KK}, 
Einstein--Gauss--Bonnet\cite{egb}, Einstein--Cartan\cite{tors} and 
nonsymmetric field\cite{ns} theories. Some dynamical, {\em i.e.},
time-dependent, solutions have also been proposed\cite{dyn}.

For the Euclidean case, there are some theorems about the nonexistence
of vacuum, {\em i.e.}, Ricci flat, wormholes. In \cite{GP}, 
it was presented a theorem stating that any asymptotically flat,
nonsingular,
Ricci flat metric in $R^4 - \{N {\rm\ \ points}\}$ is flat. 
The proof uses some topological invariants of four-dimensional
manifolds that can be expressed by means of integrals of curvature 
invariants,
{\em viz.} the signature
$\tau({\cal M})$ and the Euler characteristic $\chi({\cal M})$.
 The latter can be expressed, for a closed orientable diffentiable manifold
$\cal M$
of dimension $n=2p$ endowed with a Riemannian metric $g$,
by the Gauss-Bonnet
formula as:
\beq
\label{gb}
\int_{\cal M} \epsilon_{a_1\cdots a_n} R^{a_1 a_2}\cdots R^{a_{n-1}a_n}
 = (-1)^{p-1} 2^{2p}\pi^p p! \chi({\cal M}),
\eeq
where 
$R_a^{\ b}$ is the curvature 2-form of $g$. The signature can also
be expressed by an integral of curvature invariants. (See \cite{GP}
for references).

One can consider a four-dimensional manifold with $N+1$ asymptotically
flat regions as being topologically equivalent to $R^4$ with
$N$ points removed; one to each additional asymptotic region. Hence,
the
result of \cite{GP} rules out the existence of vacuum nonsingular 
euclidean wormhole solutions. 
This result was extended to
the non-empty case in \cite{JW}, where it is shown that there is
no nonsingular
euclidean wormhole satisfying Einstein equations with  conformal
invariant matter fields obeying  appropriate falloff conditions
in the asymptotic regions.

For the lorentzian case, there are no equivalent theorems in the
literature. We remind
a long standing theorem  
due to Lichnerowicz\cite{lich},  
which states
that any stationary, complete, asymptotically flat, and Ricci-flat 
lorentzian metric in $R^4$ is flat. However, this theorem 
 cannot be extended
 to the case of many asymptotically flat regions 
($R^4 - \{N {\rm\ \ points}\}$). 
The purpose of the present work is to contribute to fill this gap
with the following theorem: 
\begin{description}
\item[]{\em  Any asymptotically flat, static,
nonsingular, Ricci flat lorentzian 
metric in $R^4 - \{N {\rm\  points}\}$ is flat. } 
\end{description}
We will show that for the non-empty
case some nonexistence theorems can also be formulated.
The  approach to prove our main result, based on the Gauss-Bonnet formula,
will be similar to the used in \cite{GP}.
Although Gauss-Bonnet theorem is rather subtle for other 
signatures\cite{alty}, we can use Chern's 
intrinsic proof\cite{chern} of the Gauss-Bonnet formula (\ref{gb}) with
minor modifications. 

Let us now briefly review some points of Chern's proof with relevance to
our purposes. In \cite{chern}, Chern  showed that $\Omega$
can be written as $\Omega = d\Pi$ for a suitable $(n-1)$-form $\Pi$. 
We will consider here this problem for the case of an 
open four-dimensional
manifolds with a lorentzian complete metric.

Be $V$ a continuous unit  timelike
vector field  $(V_aV^a = -1)$.  Note that contrary to the
case of $\cal M$ closed, here there is no topological obstruction 
to the global
existence of such a vector field. Let us introduce the vector-valued
1-form
\beq
\label{e3}
\theta^a = DV^a = dV^a + \omega_b^aV^b,
\eeq
where $\omega_b^a$ is the Levi-Civita connection 1-form. Due to that
$V^a V_a = -1 $, the 1-forms $\theta^a$ are linearly dependent, indeed
$V_a\theta^a=0$. From (\ref{e3}) we have
\beq
\label{e4}
d\theta^a = \theta^b\wedge \omega_b^a + R_b^{\ a} V^b.
\eeq

Now, consider the following 3-forms
\begin{eqnarray}
\phi_0 &=& \epsilon_{abcd} V^a\wedge \theta^b\wedge \theta^c 
                              \wedge \theta^d, \nonumber \\
\phi_1 &=& \epsilon_{abcd}  V^a\wedge \theta^b\wedge R^{cd}.
\end{eqnarray}
Using the linearly dependence of the $\theta^a$, the Bianchi's identities,
(\ref{e3}), and (\ref{e4}) we get
\begin{eqnarray}
\label{e5}
d\phi_0 &-& 3\epsilon_{abcd}V^a\wedge V_f \wedge R^{fb}\wedge \theta^c 
        \wedge \theta^d = {\rm terms\ prop.\ to\ }\omega_b^a,\nonumber\\
d\phi_1 &-& \epsilon_{abcd}\left( 
\theta^a\wedge\theta^b\wedge R^{cd} + V^a\wedge V_f \wedge R^{fb}
                                                    \wedge R^{cd}
\right) = {\rm terms\ prop.\ to\ }\omega_b^a.
\end{eqnarray}
The left-handed sides of (\ref{e5}) are clearly intrinsic expressions, and, 
by choosing normal coordinates
around an arbitrary point,
we
can easily show that they indeed vanish. 

We can cast (\ref{e5}) in a more convenient form by using
 the generalized Kronecker delta
\begin{equation}
\label{kro}
\delta^{a_1 \cdots a_n}_{b_1 \cdots b_n} =
\left|
\begin{array}{ccc}
\delta^{a_1}_{b_1} & \cdots & \delta^{a_1}_{b_n} \\
\vdots & { }   & \vdots \\
\delta^{a_n}_{b_1} & \cdots & \delta^{a_n}_{b_n}
\end{array}
\right|.
\end{equation}
For $n>4$, $\delta^{a_1 \cdots a_n}_{b_1 \cdots b_n}$ vanishes
identically. In particular, one has
\beq
\label{e6}
\delta^{a'b'c'd'f'}_{a\:b\:c\:d\:f} = 
\delta_a^{f'} \delta_{b\:c\:d\:f}^{a'b'c'd'} - 
\delta_b^{f'} \delta_{a\:c\:d\:f}^{a'b'c'd'} +
\delta_c^{f'} \delta_{a\:b\:d\:f}^{a'b'c'd'} - 
\delta_d^{f'} \delta_{a\:b\:c\:f}^{a'b'c'd'} + 
\delta_f^{f'} \delta_{a\:b\:c\:d}^{a'b'c'd'} = 0
\eeq
Using that $V_aV^a = -1$, the linearly dependence of $\theta^a$,
 and (\ref{e6}) we obtain
\begin{eqnarray}
\label{e7}
\epsilon_{abcd}V^a\wedge V_f \wedge R^{fb}\wedge \theta^c 
        \wedge \theta^d &=& -\frac{1}{2} \epsilon_{abcd} 
                        R^{ab}\wedge\theta^c\wedge\theta^d, \nonumber \\  
\epsilon_{abcd}V^a\wedge V_f \wedge R^{fb}\wedge R^{cd }
                        &=& -\frac{1}{4} \epsilon_{abcd} R^{ab}\wedge R^{cd}.
\end{eqnarray}
{}From (\ref{e7}) we finally get
$\Omega = \epsilon_{abcd} R^{ab}\wedge R^{cd} = d\Pi,$ where 
\beq
\label{e8}
\Pi = -4 \left(\phi_1 - \frac{2}{3}\phi_0 \right) = -4
\epsilon_{abcd}V^a\wedge\theta^b\wedge \left(
R^{cd} - \frac{2}{3} \theta^c\wedge\theta^d \right).
\eeq
The expression (\ref{e8}) is essentially the result of Chern. 
The difference
is that in the case of closed manifolds, $\Pi$ is not defined everywhere.
In general,  continuous 
vector fields have some singular points in a closed manifold, 
and thus we cannot construct a vector field with $V_a V^a\ne 0 $
everywhere. As already said, in the present case there is no
topological obstruction to the existence of a nowhere vanishing 
vector field.

We can now prove our result. 
For static, Ricci-flat lorentzian metrics
$\Omega$ is non negative. To see it,
first note that with $R_{ab}=0$ one has
\beq
\Omega = \frac{1}{6} \left( R_{abcd}R^{abcd} 
-4 R_{ab}R^{ab} + R^2\right)\nu 
= \frac{1}{6} R_{abcd}R^{abcd} \nu,
\eeq
where $\nu$ is the standard volume form in $R^4$. 
Consider now normal coordinates
around an arbitrary point $P$. We have
\beq
\label{e121}
\Omega_P = \frac{\nu_P}{6}  \left(
 -4\sum_{\alpha\beta\gamma} \left(R_{1\alpha\beta\gamma}\right)_P^2
 +4\sum_{\alpha\beta} \left(R_{1\alpha 1\beta}\right)_P^2
 +\sum_{\alpha\beta\gamma\delta} \left(R_{\alpha\beta\gamma\delta}\right)_P^2
\right).
\eeq
Hereafter, Roman and Greek indices run, respectively, 
over $\{1,2,3,4\}$ and $\{2,3,4\}$. Due to the assumption
of a nonsingular $g$, we have $\Omega < \infty$. 
The hypothesis that $g$ is
static assures  that there is always a coordinate system 
$\{x^a\}$ for which
\beq
\label{e14}
g_{ab} = \left( 
\begin{array}{cc}
-\Delta & 0 \\
0 & h_{\alpha\beta}
\end{array}
\right),
\eeq
where $\Delta>0$ and $g_{ab}$ does not depend on $x^1$. One can check that
for metrics of the type (\ref{e14}), 
$R_{1\alpha\beta\gamma}=0$, and consequently $\Omega \ge 0$.  Moreover,
the equality holds only if $R_{abcd}=0$. Thus, in the case of
$\int_{\cal X} \Omega = 0$,  we have that
$R_{abcd}=0$ in ${\cal X}$. This will be the strategy of our proof.
The assumption of an asymptotically flat 
metric guarantees that  for each
asymptotic region one has
\begin{eqnarray}
\label{e715}
\lim_{r\rightarrow\infty} \Delta - 1 &=&
O^\infty\left( \frac{1}{ r^{m+\varepsilon}}\right), \nonumber \\
\lim_{r\rightarrow\infty} h_{\alpha\beta} - \delta_{\alpha\beta} &=& 
O^\infty\left( \frac{1}{ r^{m+\varepsilon}}\right).
\end{eqnarray}
for some $m \in \{0,1,2,\dots\}$ and 
$0<\varepsilon\le 1$, where  
$r^2 = x^\alpha x^\beta \delta_{\alpha\beta}$. 
If $F=O^\infty(f)$ it means that
$F = O(|f|)$, $F' = O(|f'|)$, and so on.
We can construct 
the 3-form $\Pi$ starting from the timelike unit vector
$V=\frac{1}{\sqrt{\Delta}}\frac{\partial}{\partial x^1}$. From (\ref{e715}),
we obtain the following expression, valid for each
asymptotic region, 
\beq
\label{pe}
\lim_{r\rightarrow\infty}\Pi = 
O^\infty\left( \frac{1}{ r^{m+3+\varepsilon}}\right).
\eeq
Integrating $\Omega$ 
over ${\cal M} = R^4 -\{N{\rm\ points}\}\approx S^4 -\{(N+1)\rm\ points\}$
and with the assumption of a nonsingular $g$
one has
\beq
\label{int}
\int_{{\cal M}}\Omega = \sum_{i=1}^{N+1}\int_{\partial_i{\cal M}} \Pi,
\eeq
where the boundaries $\partial_i\cal M$ correspond to the asymptotic
regions. For each of these regions, with the asymptotic 
conditions (\ref{e715}), we have
\beq
\label{e13}
\lim_{r\rightarrow\infty} \mu\left( 
\partial_i {\cal M}\right) = O^\infty\left( r^3\right), 
\eeq
where $\mu\left( \partial_i{\cal M}\right)$  denotes the measure 
of the boundary $\partial_i{\cal M}$. From (\ref{pe}) 
and (\ref{e13}) we have finally
that the right handed side of (\ref{int}) vanishes, 
establishing our result.$\Box$

It is shown in \cite{JW} that, for the euclidean case, sometimes
the matter equations
themselves can be used to rule out non-vacuum wormholes. The same
arguments can be applied here for some static matter fields. Let us take,
for instance, a real scalar field $\phi(x)$ 
with a potential obeying $\phi V'(\phi)\ge 0$. The hypothesis of
a nonsingular $g$ requires that $\phi$ and $\partial_a\phi$ 
be smooth and bounded on $\cal M$. The corresponding
equation in this case is
\beq
\label{scal}
D_a D^a\phi = V'(\phi).
\eeq
Multiplying by $\phi$ and integrating over $\cal M$ one obtains
\beq
\label{ia}
\int_{\cal M} 
\left( g^{ab}(\partial_a\phi)(\partial_b\phi) 
+ \phi V'(\phi) \right)\, d{\rm vol}
- \sum_{i=1}^{N+1}\int_{\partial_i \cal M} \phi 
\partial_a\phi\, d\Sigma^a = 0.
\eeq
With the assumption that $\phi$ obeys an asymptotic condition like 
\beq
\label{iai}
\lim_{r\rightarrow\infty} \phi =  O^\infty\left(
\frac{1}{m+\varepsilon}\right),
\eeq
the boundary
terms in  (\ref{ia}) vanish.
This assumption requires that $V'(\phi) = 0$ for $\phi=0$.
The first term  in (\ref{ia}) is, 
for the static case, nonnegative, implying that
$\phi$ must vanish in $\cal M$ if $\phi V'(\phi)\ge 0$. 
We make here the same remark done for the euclidean case\cite{JW}:
if $V'(\phi)=0$ for some $\phi\ne 0$, and $\phi$ assumes to a nonzero
value in some of the asymptotic regions, the boundary term in (\ref{ia})
may not vanish in general. A nonexistence result holds also for 
conformally coupled massless fields. In this case, we have
\beq
D_aD^a\phi - \frac{R}{6}\phi =0. 
\eeq
However, Einstein equations imply in this case that $R=0$, and we
get in fact a particular case of (\ref{ia}).

Another example of nonexistence of non-empty wormhole solutions
is the case of massless fermions. In this case, the matter
equations are
\beq
i\partial_a\gamma^a\psi = 0.
\eeq 
Applying the Dirac operator $i\partial_a\gamma^a$ again one gets
a conformally coupled Klein-Gordon equation for each component
of $\psi$
\beq
\label{p}
\left( D_a D^a - \frac{1}{6}R \right)\psi = 0.
\eeq
Einstein equations also imply that $R=0$ in this case. Also,
we can redefine $\psi$ in order to have only real components.
As in the previous cases,  provided that  $\psi$ obeys
appropriate falloff conditions in the asymptotic regions, we
conclude that there is no wormhole solution with massless fermions.

We finish noticing that, with the same procedure use here, we can
show that in $R^4 -\{ N\rm\ points\}$, 
any metric of signature $(-,-,+,+)$, 
Ricci-flat, asymptotically flat,
static simultaneously with 
respect to two linearly independent time-like Killing vectors,  
is flat.

\acknowledgements

The author is grateful to DAAD, CNPq and FAPESP for the financial support,
and to Prof. H. Kleinert and Dr. A. Pelster for the warm hospitality at
the Freie Universit\"at Berlin.

\end{document}